\documentclass{aa}  
\usepackage{graphicx}
\usepackage{txfonts}
\begin{document}

   \title{Solar Active Region Electric Currents Before and During Eruptive Flares}

 %  \subtitle{}

   \author{
          %\inst{1}
	 B. Schmieder \inst{1}
	  \and
	  G.Aulanier\inst{1}
	}

   \institute{LESIA, Observatoire de Paris, PSL Research University, CNRS, Sorbonne Universit\'es, UPMC Univ. Paris 06, Univ. Paris Diderot, Sorbonne Paris Cit\'e, 5 place Jules Janssen, F-92195 Meudon, France
   \email{brigitte.schmieder@obspm.fr}
	       }

   \date{Received ...; accepted ...}

% \abstract{}{}{}{}{} 
% 5 {} token are mandatory
 
  \abstract
   { The chapter  "Solar Active Region Electric Currents Before and During Eruptive Flares" is a  discussion on  electric currents in the pre-eruption state and
in the course of eruptions of solar magnetic structures, using information
from solar observations, nonlinear force-free  field extrapolations
relying on these observations, and three-dimensional magnetohydrodynamic
(MHD) models. The discussion addresses the issue of neutralized vs.
non-neutralized currents in active regions  and concludes that MHD models are able to explain non-neutralized currents  
in active regions by the existence of strong  magnetic shear along the polarity inversion lines, {  thus confirming previous observations that already { contained} this result.}
The models  have also captured the essence of the behavior of electric
currents in active regions during solar eruptions, predicting current-density increases and
decreases inside flare ribbons and in the interior of expanding flux ropes respectively.
The observed photospheric current density maps, inferred from vector magnetic field observations, exhibit  similar whirling  ribbon patterns  to the MHD model results, that are  interpreted as the  signatures of flux ropes and  of quasi-separatrix layers (QSLs) between the magnetic systems in active regions.
Enhancement of the total current in these QSLs during the eruptions { and decreasing current densities at the footpoint of erupting flux ropes,} has been confirmed in the observations.
    }
   {
   } 
   {
   } 
   {
     }
     {  } 

   \maketitle
%
%________________________________________________________________

\section{INTRODUCTION}

%Figure 2-1 
 \begin{figure*}
   \centering
    \includegraphics[width=8.0cm]{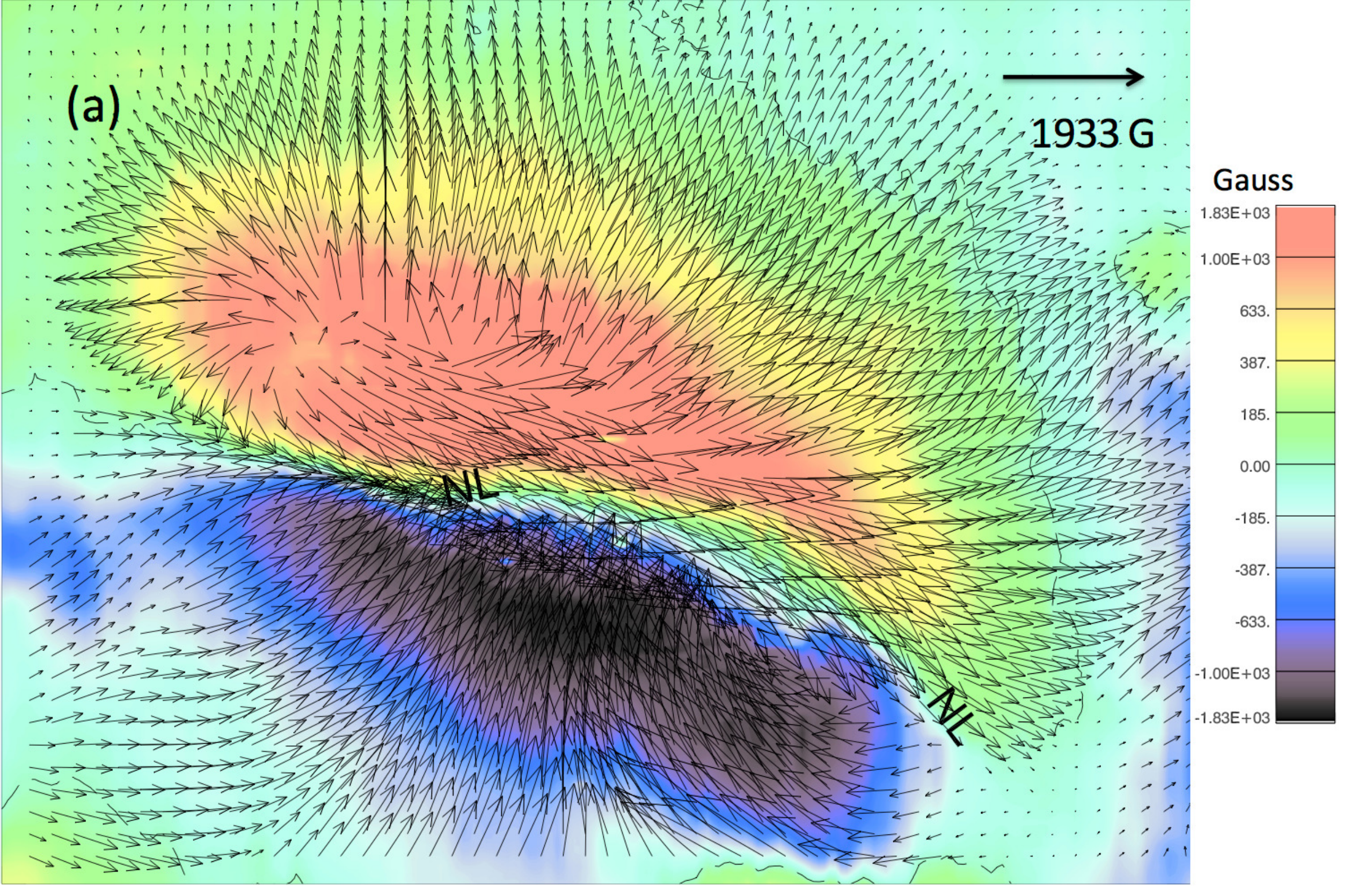}
    \includegraphics[width=5.0cm]{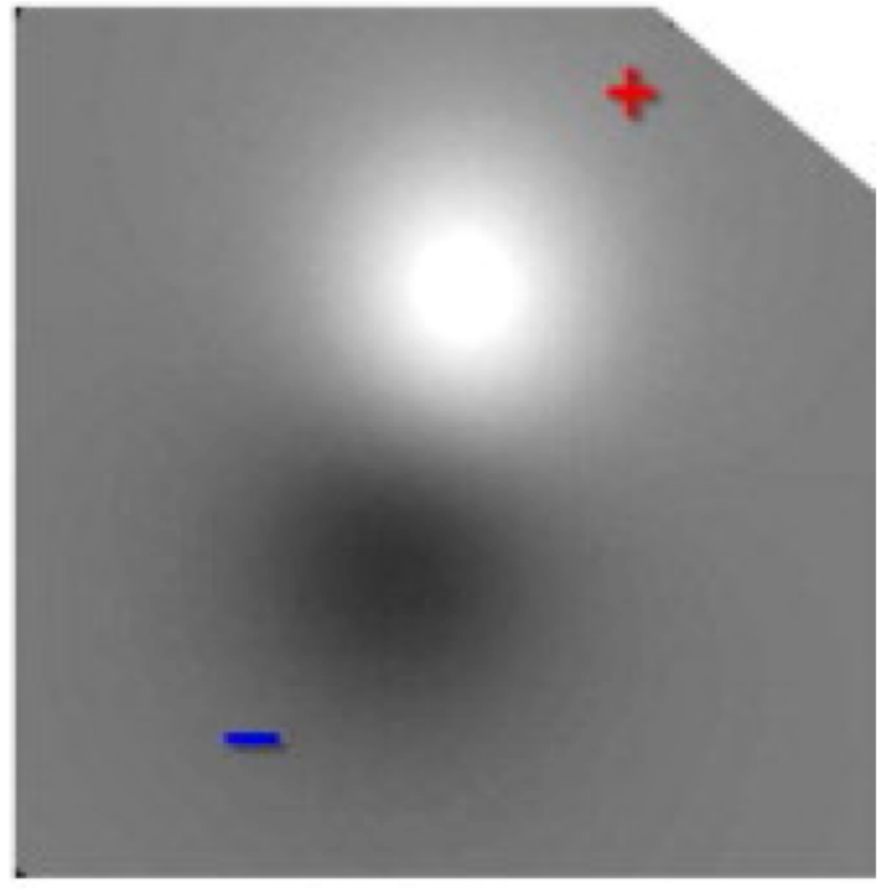}
     \includegraphics[width=7.8cm]{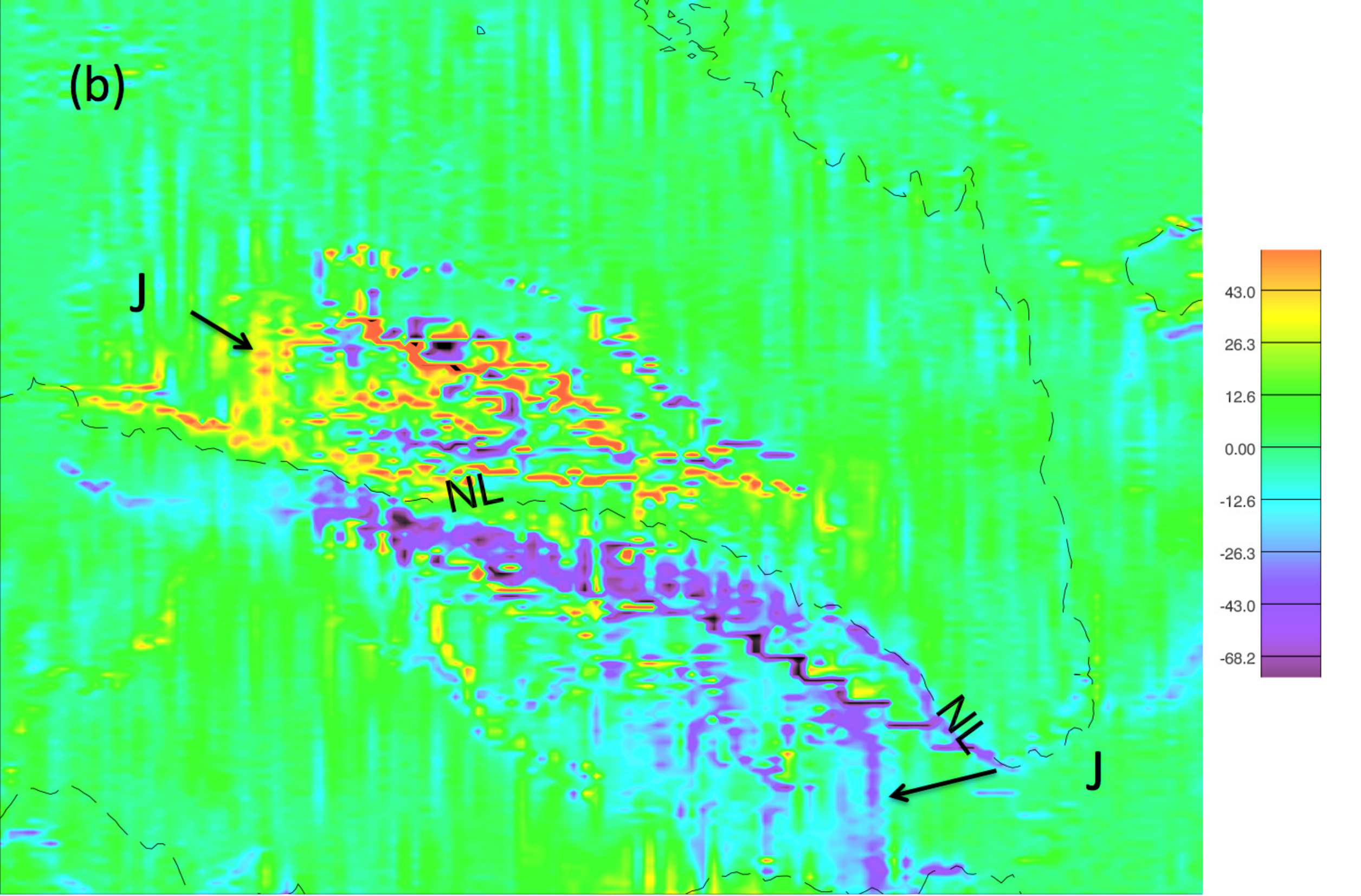}
     \includegraphics[width=5.0cm]{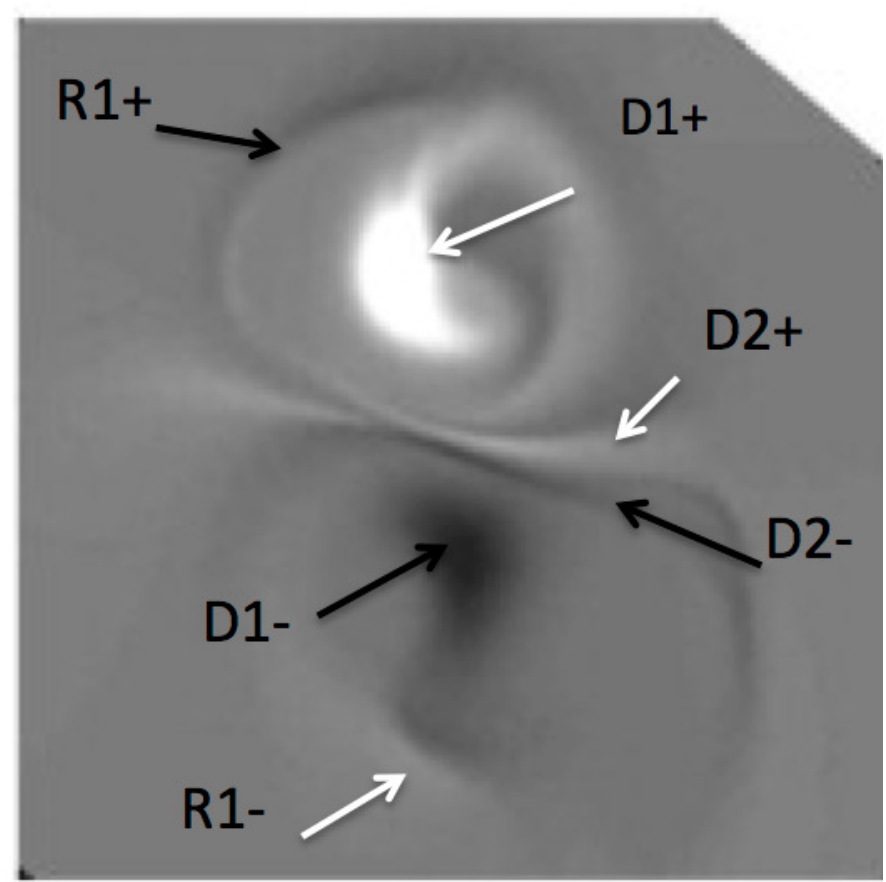}
           \caption{Vector magnetogram and vertical component of the electric  current density map  in the  emerging active region  {  AR 10808 } with positive helicity observed with the THEMIS vector magnetograph on September 13 2005 (Courtesy of Bommier). (a) $B_z$  map with  two elongated tongues symmetrical to the neutral line (NL or PIL) where  the arrows represent the horizontal  field (red-yellow represents the positive polarity, blue-violet the negative polarity between +/- 1800 Gauss).   (b) Vertical component of the electric current  density in  units of  mA    m$^{-2}$; yellow represents the positive  current while violet the negative current. The  letters J and arrows point out  the J-shaped current ribbons.  ({\it Right  panels}):  Initial  $B_z$   in a bipole and  vertical electric current  density  after  applying a positive twist in the OHM  simulation (adapted from \citet{Schmieder2012}, see Section 2.2). White  areas correspond to  positive $B_z$ and  $J_z$, dark-grey areas to  negative  $B_z$ and  $J_z$. 
           D indicates direct currents, R  return currents. Their sign is defined by the magnetic helicity convention shown in Table1.
           }
   \label{THEMIS}
   \end{figure*}

   The occurrence of finite electric currents in the Sun's corona 
is a fundamental requirement for the existence of an
active phenomena, such as solar flares and CMEs. Several reasons 
can be put forward. 
In the dilute corona, the plasma beta ($ \beta$) is much lower than unity. 
So the structure of the coronal plasma is dominated by  the magnetic 
field, and the magnetic energy is the main reservoir for all dynamics. 
Also, active phenomena  occur on time-scales far shorter 
than those on which the dense photosphere  evolves. So the 
distribution of magnetic flux crossing the photosphere hardly changes 
during solar transients. Given the solenoidal nature of the 
magnetic field, the potential (i.e. current-free) field corresponds 
to the lowest magnetic energy state for a  fixed photospheric flux 
distribution. 
  However, slow photospheric motions can induce twist and stress of the magnetic field lines and lead to  a non potential magnetic field, which carries electric currents.
 So current-carrying magnetic fields constitute the 
source of energy for active phenomena. 
Also beyond this global property, electric currents play several 
key local roles in magnetohydrodynamics (MHD). The most significant 
examples are their presence in the Lorentz force, as well as in several terms of the generalized Ohm's law, including the usual resistive 
term and the Hall electric field which are both involved in magnetic 
reconnection within current sheets. 

   In spite of their importance,  few studies have been realized on
solar electric currents over several decades and up to very recently. The reason is probably  theoretical,  and observational.

Theoretically, several historical models  had initially
 considered the electric current to be an input parameter. For example,
``circuit models'' for solar flares 
prescribed the total current $I$ as it is in laboratory experiments
 \citep[see e.g.][]{Alfven67,Spicer1982}. 
 More
generally, one approach to plasma physics that is often used in the
magnetospheric community \citep[and is advocated by e.g.][]{Melrose95debate,
Heikkila97} is to make every calculation with the equations that use
the electric field $\vec{E}$ and the current density $\vec{J}$ as
variables. This is the so-called ``$\vec{E};\vec{J}$ paradigm''. But both   circuit models and the $\vec{E};\vec{J}$
paradigm have been  criticized throughout the years \citep[see][]{Parker96enddebate,Parker96vBEj,Parker01vBEj}. In this line,
E.N. Parker argued that they should be substituted by the ``$\vec{B};
\vec{v}$ paradigm'' in MHD. In the latter, the current densities $\vec{J}$
merely result from the generation of  a finite curl of the magnetic field $\vec{B}$
through Amp\`ere's equation, as a direct result of plasma flows $\vec{v}$
through the ideal induction equation. So, the current is not a prime
variable, since it does not even appear in the governing equations of MHD. For this reason, electric currents have   been underlooked for years.

Observationally, the photospheric magnetic field  vector in solar active regions  provides
the vertical component of curl $\vec{B}$.  However the vector magnetic field can be measured only in the photosphere, and  thus only   the vertical component of the current density $J_z$ passing through the photosphere to the corona can be  identified. 
An idealized active region (AR) consists of two sunspots represented  {  by } a bipole with two opposite polarities linked by{  a flux tube (called a flux rope when it is twisted)}  in the corona.
{ Since $\beta<1$ in sunspots, the { AR-scale magnetic-field}  should be close to force free not only in the corona but also in the photosphere. 
Thus electric current density of   both signs should be   measured in sunspots indicating  a flow of currents in both directions { in the  large scale coronal flux tube}.}
Most (if not all) vector magnetic field measurements 
used  are  quite sparse as well as relatively unreliable outside sunspots 
and their close vicinity. So photospheric electric-current   densities    are rather 
difficult to measure using the transverse magnetic field, which  was and still is,  commonly very noisy.
In  past observations only one sign of electric current was detected in each polarity. A full neutralization of currents in an active region implies that the  total net current   is   zero in  each polarity of a bipole.
Therefore, results such as  net currents  in active regions  were regarded as uncertain \citep[as reported by e.g.][]{Gary87,Hagyard1988,Wilkinson92,Leka1996,Leka99}.
 With the development of 
new-generation and dedicated ground-based telescopes (e.g. THEMIS in Canary Islands) and space-borne 
(e.g. Hinode, {\it Solar Dynamic Observatory} SDO) missions,  more complicated electric current patterns have been detected in each polarity of  ARs.

Thus  new MHD simulations  have been developed  during  recent  years.   Some  correspondence has  been found with the    observations. 
MHD simulations   based on the existence of  twisted  flux ropes show that   current densities of both signs exist.   A  twist of finite radius theoretically implies the  occurrence of  a sheath of currents at the edge of the twisted flux tube, which flows in  an opposite direction to the direct currents. These are called the return currents. 
A twisted flux tube of left-handed twist has a negative helicity and a direct electric current that flows anti-parallel to the magnetic field (and vice versa) \citep{Demoulin2007,Demoulin2009}.  
{ { The modern conventional definition in MHD simulations  which holds for  AR large-scale   quasi force-free coronal-field} is summarized in Table 1. Actually, this definition is consistent with the inference of direct current (i.e. with their dominant sign, as mentioned above) in past observations.   
%There is neutralization if  the direct  current is  equal  to the return current, which  is a rare situation.

 {  Indeed,  observational    studies by \citet{Wheatland2000,Falconer2002,Ravindra2011,Georgoulis2012,Schmieder2012} showed  new  and frequent  evidences of net current in solar active regions. }These relatively recent observations have revived the old debate from the "90s" about  
% to prove that net current exist. A few of them have addressed the old problem debated in the ''90s"  about 
  the neutralization,  or not,   of  electric currents in active regions \citep{Melrose1991,Parker96enddebate}. 
{  In this perspective the MHD simulations developed  by \citet{Torok2014,Dalmasse2015} }
have explained  how and why 
 unneutralized current densities  occur in active regions when   magnetic shear is present
% shear motions (i.e. flows)
   along   polarity inversion lines (PIL). }

In parallel, progress on the { description of the } magnetic topology of  active regions has contributed significantly to the  interpretation of  the current density patterns observed before and during eruptions. 
The field-aligned currents in line-tied  force-free coronal structures,
imply a direct  continuity between the coronal and the photospheric  currents, especially { above}  sunspots where $\beta <1$. 
 Because of magnetic flux conservation along the flux tube,
 one can  expect to see   the footprint of the coronal  current  densities    in the photosphere. 
In other words,  the photospheric current pattern in sunspots  should show a cross-section of the  electric    current density  distribution inside the flux tube.
In addition to  twisted flux tubes with their volume currents, other magnetic structures also  exist in the corona and connect down to the photosphere. Among such structures, one can find open-closed boundaries { that separate  closed magnetic field above active regions from  open field in their surroundings}, and  separatrices between different flux systems,  or  quasi-separatrix layers  (QSLs), which  are very narrow  volumes 
across which field lines quickly change connectivity. 
 In these thin  volumes,   electric  current sheet can develop along their length and eventually trigger  reconnection of the magnetic field lines,  leading to flares \citep{Demoulin1996a}.
Therefore during eruptions  QSLs become very important  structures.
 The footprints in the  photosphere  of these structures are  thin lanes of QSls, { which appear as  long and  narrow hooks. These structures   reveal} the evolution of the electric current densities  before and during eruptions,  observationally and  in MHD models \citep{Janvier2014}.

\begin{table}
\begin{tabular}{ccc}
\hline
 Magnetic helicity &  direct current & return current \\
 \hline
$ H_B$  $> 0$ & $J_Z$  $B_Z$ $>0 $ & $J_Z$ $B_Z$ $<0 $\\
$H_B$  $< 0$ & $J_Z$ $B_Z$ $<0 $ & $J_Z$ $B_Z$ $>0 $\\
\hline
\end{tabular}
\caption{Electric current  density signs for positive and negative magnetic  helicity {    at active region scale.}}
\end{table}

This chapter addresses some of these issues, and describes new  
publications with  recent findings and interpretations resulting from coupling current measurements 
and MHD models.  The next  section is focused on the evidence of  pre-eruption currents in active regions (Section 2), and the following section concerns the 
evolution of the currents 
during  eruptions (Section 3). We conclude in Section 4.

\section{PRE-ERUPTION CURRENTS IN ACTIVE REGION}

\subsection{Observations of Direct and Return Current }

The photospheric current distribution is derived from measurements of the vector magnetic field  using Amp`{e}re's law.
%(J proportional to  curl B).
 The vector magnetic  field  is measured by using the Zeeman effect which is sensitive mainly in photospheric lines.
Hence only the vertical component  of electric current density  $J_z$ can be readily calculated in the photosphere. 
The first  published calculations  of $J_z$ were presented  by \citet{Severny1964} and \citet{Moreton1968}  (see the thesis of Harvey 1969) and later 
by \citet{Hagyard1988,Wilkinson92}, and \citet{Canfield1993}. The $J_z$ observations  showed principally  two areas of opposite sign  electric current density  partly overlying  the leading and following  polarities of  active regions (AR). Electric currents in magnetically  isolated regions  flow  from one polarity to the other polarity,  and correspond to direct currents. This was the basis of flare models with unneutralized active regions.

 Due to recent improvements in vector magnetographs the measurements of electric  current density  are more reliable and  
  now    direct  and return  electric currents are  detectable.
         Recent  measurements indicate  commonly that the return currents  in total  are much smaller than the direct currents and a net current still   exists. From recent observations using high  spatial  resolution vector magnetograms ({\it the Helioseismic and Magnetic Imager}  HMI aboard SDO,  {\it the Spectro Polarimeter} SP aboard Hinode) and  ground based  telescopes  (e.g. THEMIS)  active region current maps   are routinely    obtained.   ARs with  neutralized  currents and  ARs with  unneutralized currents  have been  both observed \citep{Wheatland2000,Metcalf2006,Ravindra2011,Georgoulis2012,Gosain2014,Schmieder2015,Cheng2016,Zhao2016}.   
     
We select one example of  an AR with  electric current density observations obtained with THEMIS.  From the retrieved Stokes parameters {\it IQUV} of the vector magnetogram  the three magnetic field components were obtained  by using the Milne-Eddington inversion code UNNOFIT \citep{Bommier2007}.   The active  region was a new born active region (AR) \citep{Li2007,Canou2009,Bommier2013,Schmieder2012,Schmieder2015}.
The photospheric magnetic field  of the  active region presented two elongated tongues of opposite polarities (Figure \ref{THEMIS} a).  
The photospheric  magnetic field  vector direction (see the arrows in Figure \ref{THEMIS}a) suggests that the twist of the emerging flux tube is right handed i.e. has   a positive helicity.
It has been shown that such a    tongue-shape  pattern in the photospheric  magnetograms  indicates that the emergence of  the  flux tube is  not completed
\citep{Chandra2009,Luoni2011}. In fact, the tongue elongation pattern indicates the existence of an azimuthal  component of the magnetic field  around the axis of the Omega-shaped  coronal flux tube, and therefore  the sign of the  twist of  the emerging flux tube can be deduced  (see Section  2.2).  
 The current pattern shows J-shaped current ribbons  (Figure \ref{THEMIS}b).  
    The direct current occupies  two J-shaped areas with   two elongated lanes  in the AR center  along the PIL.  
 The  direct current density   $J_z$  has a positive sign over the positive magnetic polarity and vice versa   according to the positive helicity  of the active region (Table 1).
  { At low resolution} return currents  are observed at the periphery of each sunspot  as much weaker current densities   and narrower current lanes.
   { At high spatial resolution it should be still noted that the fibril nature of sunspot penumbrae there leads to mixed strong direct and return currents at small scale \citep{Venkat2009}.}
   % mixed with the direct currents { due to the presence of fibrils of different inclination in the penumbra}. 
   The observations of such return currents are  at the limit of what can be  identified on the edge of and away from sunspots
   because the photosphere is  there far from force free, {  because at small scale the magnetic field is subject to granular motions.}
 So  we are only concentrating here on  current in sunspots where $B$ is  sufficiently strong  ($\beta<1$).

These observations  have been compared with theoretical results obtained by a  MHD model created from a bipolar potential field by photospheric vortex flows using the OHM   code \citep{Aulanier2012} (see Section 2.3).  The $ \beta$ = 0 simulation
performed with the OHM code reproduced the evolution
of an initially torus-unstable flux rope    (Figure \ref{THEMIS} right column).  

    The global pattern of the  photospheric  { electric current density}  shows  intense  direct currents in the polarities  surrounded by return currents with a J-shape,  similar to     the observations.       The dominant direct  current  in the strong fields (encircled by  hooks)  are (D1+ and D1-).    The hooks primarily have direct currents, surrounded by weaker return current  densities   (R1+ and R1-).  D2+ and D2- are   current density lanes in the central part of the active region.

The presence of  these relatively narrow current density ribbons  in the photosphere
 can be explained as follows.
  The connectivity domains are bordered by  the quasi-separatrix layers (QSLs) where large magnetic field distortion  could exist. 
 Between the flux rope and the environment,  coronal current density  layers can be  formed  during the pre-eruptive phase. 
 Along the  QSLs, any small perturbation can  induce an  increase of the currents in  these narrow  layers   which  are rooted in the photosphere \citep{Demoulin1996a}.  Therefore analyzing the electric current  density maps in the photosphere  can inform on the existence and the geometry of QSLs, and hence on the location of a flux rope.
In particular, the hook-shape extremities of the  current ribbons  in the current  density maps are   the signatures of  the existence of a  twisted flux rope. 
The curvature of the hook part of the J depends on the twist of the flux rope \citep{Demoulin1996b}.  
{ The review of \citet{Gibson2006} also provided  the evolutionary picture accompanying the flux-rope formation with hook structures.}
 
%Figure 1-2
 \begin{figure*}
   \centering
      \includegraphics[width=\hsize]{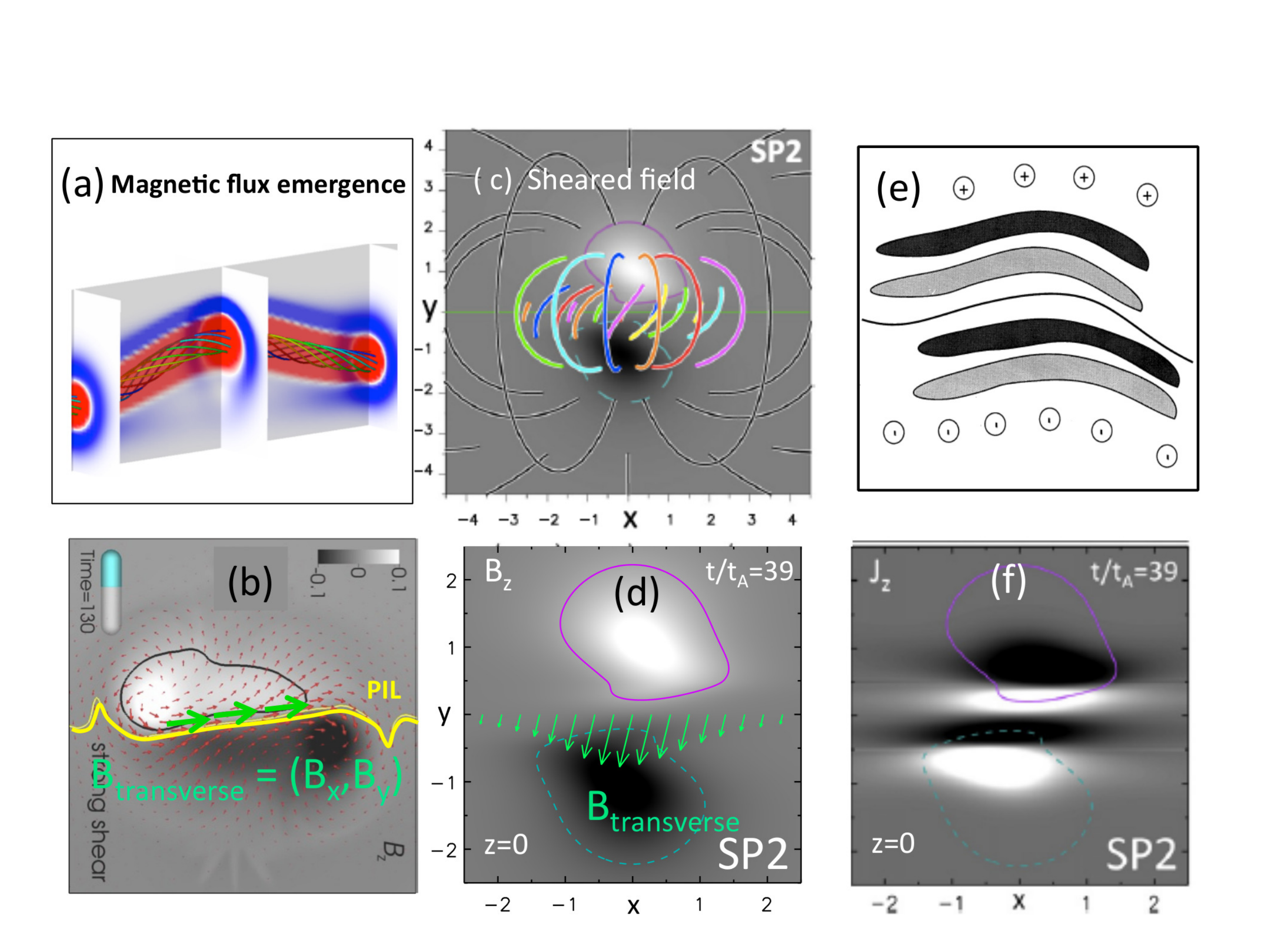}
   \caption{MHD simulations of magnetic flux emergence.  Panel (a) a model of  emerging flux rope from \citet{Leake2013}. Panels (b, c, d, f) results from the simulations of \citet{Dalmasse2015}
showing in (b, c and d) the magnetic field $B_z$ for different configurations of shear,  the  green arrows represent the transverse field  between the two polarities. In Panel (f) the   photospheric current density map $J_z$  in the configuration shown in (c) and (d)  with partially unneutralized  currents to be compared  with panel (e) showing  the predicted sketch of \citet{Melrose1991}  for  the  idealized case of neutralization. Black  areas display direct  current and  white areas return current on the positive polarity and the opposite on the negative polarity  according to the   negative sign of the helicity  (left hand side twist) (see Table1 and  Section 2.2).
}
  \label{kevin}
   \end{figure*}
   
\subsection{Current Neutralization in Active Regions: History}

{ In the previous subsection we have reported on how observations, as well as one numerical simulation, showed the coexistence of direct and return currents at active region scales. Let us now discuss the issue of current neutralization, i.e. of when, how, and why, direct and current currents can become equal.} Before describing the most recent joint observational and MHD results, we start by presenting an historical view 
%Before starting to  discuss  the recent MHD models treating  the   existence of electric currents in active regions, we present   an historical  view 
of a  long-lasting debate about whether or not a net current can exist in an  active regions \citep{Melrose1991,Parker96enddebate}.

According to  coronal  models for solar flares  \citep{Sturrock1980},  the  energy storage  involves  a non potential component of the magnetic field, which can be achieved by a twist or shear of the coronal magnetic field. 
 \citet{Melrose1991}   proposed   { a
 reductio ad absurdum argument}.
 %  a discussion by reasoning to reduce to absurdity (ad absurdum)..  
 As a first point in his discussion,  he argued  that { in an idealized case}, if  the currents were generated by photospheric or sub-photospheric stresses after an isolated magnetic flux tube had emerged, then  the currents (directed along the flux tube axis) should be neutralized.   He pursued his reasoning   by saying that if  electric currents were  neutralized  in the corona (i.e. the sum of direct currents  should equal the sum of  return currents). In the case of  { isolated flux ropes}   each direct  current should be  surrounded by return current on each  side of the  polarity inversion line (PIL).
 %a  flux rope with small twist. 
   In the case of coherent shear,   two elongated  lanes of opposite currents along each side of the PIL should be observed. { It is interesting to see that  he predicted correctly the shape of the current density pattern  { obtained more than a decade later by MHD simulations} (see Figure \ref{kevin}e  here and  Figure 2 in \citet{Melrose1991}).  But he did not correctly predict the neutralization. The reason is that  } in his argument, he considered a  cylindrical and vertical flux tube anchored in the photosphere which has emerged completely. 
   {  Treating the legs of active-region scale coronal-flux rope as being cylindrical and  orthogonal to the photosphere was an oversimplification. Indeed new MHD models show that these legs are inclined and partially  emerged  from the solar interior \citep{Leake2013,Torok2014,Dalmasse2015}.}
   
%    A model with a flux tube with no curvature} is not a correct model for flux emergence as new  MHD models  have demonstrated }.  
In a second point of his reasoning, Melrose  argued that neutralized currents were not measured (see \citet{Hagyard1988} and  Section 2.3).
With these  observational considerations, \citet{Melrose1991}  concluded  that his  basic assumption  of neutralization of  currents should   not be  valid.
  and he declared that the currents are unneutralized. {  This was a good conclusion,  however  this raised the question about }  where do the currents close. 
\citet{Spicer1982} proposed an idea 
 in the frame of  a flare model based on an electric circuit. In that case  the electric circuit was  closed in the photosphere by cross-field  horizontal  currents. Nevertheless, this   was not the solution proposed by \citet{Melrose1991},  who proposed instead  
  that the electric current should close under the solar surface,   deep in the solar dynamo region. 
 This idea allowed  the reconciliation  of  the observational results from this time with the  theoretical approach. 
 But  then, one   problem was  the occurrence of elongated paths of net/unneutralized currents in the  Sun's interior.  Such net currents implied  the existence of  non isolated flux tubes,  with structured azimuthal fields  existing around the current paths everywhere inside the Sun. The problem  was that these properties contradicted intuitive high- $\beta$ behavior that isolate in principle sub-photospheric flux ropes from their non-neutralized environment.

{  Melrose's physical thinking} of unneutralized currents was correct.  { But    it was impossible to see the role of magnetic shear at the PIL  with  a cylindrical geometry,   since PILs simply do not exist in such a  geometry. } Also  the poor observations available at this time  were not  { granted since their noise  hardly  allowed to measure return current densities, which could have been weaker than the direct current densities.}

All these ideas were 
%challenged
debated in a series of papers (e.g. \citet{Melrose95debate,Parker96enddebate,Parker96vBEj}).
Instead, 
{  \citet{Parker96vBEj}  
argued that the net current must be zero, but that, because of a failure to resolve magnetic fibril structure, measurements of the field would incorrectly infer a non-zero net current.}

 The lack of return current in the  past observations was  due either to  the return current strength being  below the threshold of the measurements,  or  to  the  low spatial resolution of the  former vector magnetographs  \citep{Leka1996},   or  to artifacts (e.g. Faraday rotation) as suggested by  \citet{Wilkinson92}. The  MHD simulations now consider 3D  Omega-shaped  loops that have  different properties than  the 
cylindrical flux tubes that were considered in the ''90s" for the discussion about electric currents. 

 Both   { modern observational  and theoretical }improvements  now permit  us to resolve  some  of the  problems debated between Parker and Melrose (see next Section).

%Figure  3 , 
 \begin{figure*}
   \centering
    \includegraphics[width=\hsize]{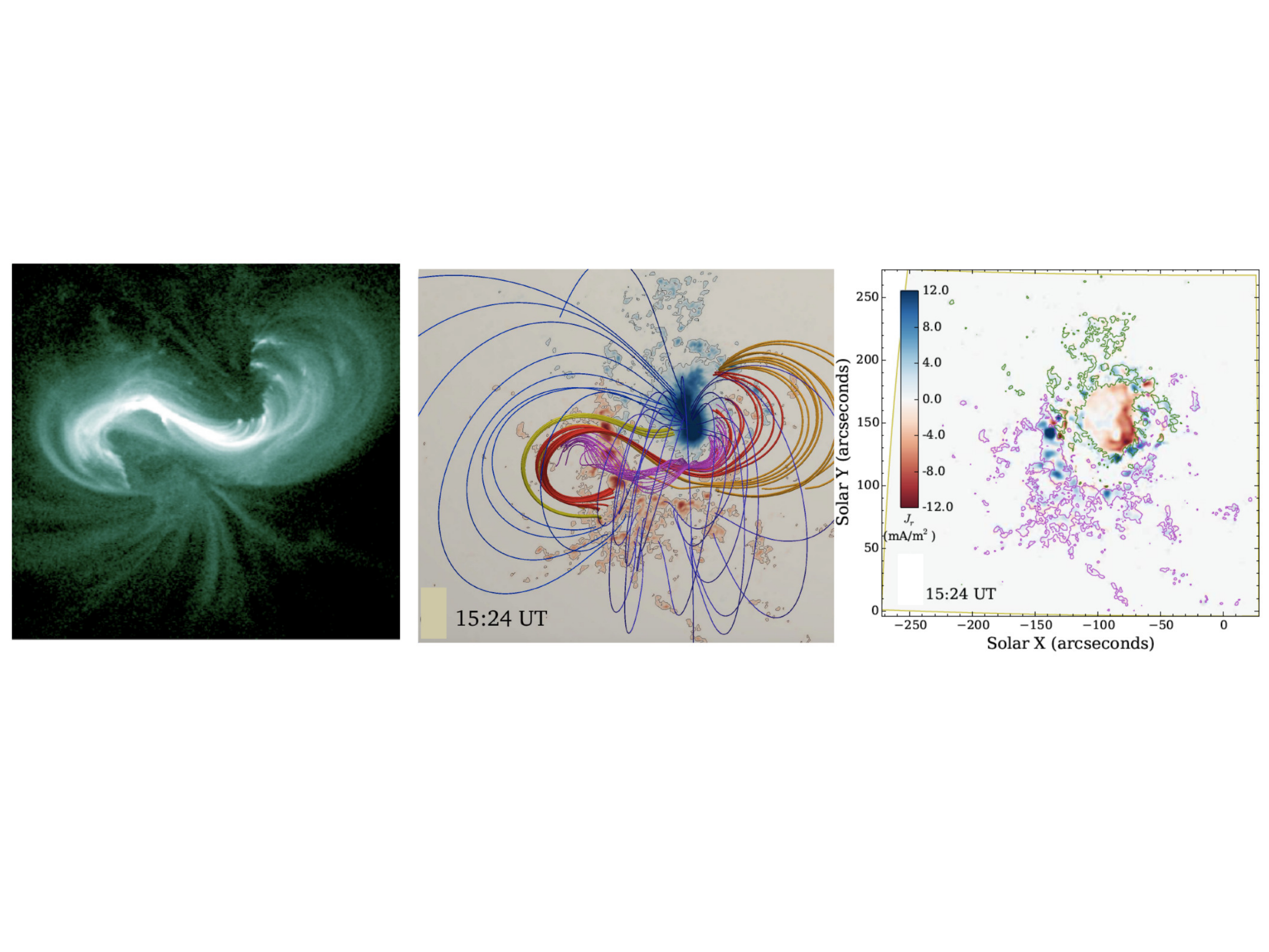}
    \vspace{-4cm}
   \caption{Sigmoidal field lines wrapping  around a flux rope in the AR 12158   witch has  a negative helicity (reverse S, left hand twist)  on September 10, 2014, two hours before a flare. The {\it left panel }shows an observation in AIA 94 \AA\ , the  {\it middle panel }shows  sigmoidal  magnetic field lines using  a NLFFF  extrapolation based on the Grad-Rubin method  \citep{Gilchrist2014}, overlying  a  $B_z$ map saturated at 2000 gauss  and the  {\it right panel} shows  direct and return electric currents  computed from HMI data after smoothing the transverse component of the observed photospheric magnetic field. The green contour outlines
leading/positive magnetic field, while the purple
contour outlines trailing/negative magnetic field   at a  field strength of  +/- 500 gauss. { Under the force-free condition,} the direct currents are red and blue in the leading positive and following negative polarity respectively { according to Table1}. The return currents are blue on the right  side of the main positive polarity { in the green contours}. The return currents (red) corresponding to the following negative  polarity are  more difficult to visualize (adapted from   \citet{Zhao2016}).}
   \label{Sept10}
   \end{figure*}

\subsection{Current Neutralization in Active Regions:  Recent MHD Models}

In the  eruptive flare models of torus-type, only a net  current  is involved \citep{vanTend1978,Molodenskii1987,Martens1989,Forbes1991,Forbes1995,Lin2000,Lin2001,Kliem2006}.   \citet{Forbes2010} showed that if  return  currents  were introduced, they could stop the MHD instability.

 Since the Melrose and Parker  discussions,  
many theoretical  models  and MHD simulations have been  proposed for flares and CMEs  \citep{Fan2003,Fan2007,Fan2010,Aulanier2010,Amari2003a,Amari2003b,Amari2014,Inoue2016}.

The recent flare and CME MHD models  have been  based on   two mechanisms:    emergence of current-carrying magnetic flux tubes through the photosphere  \citep{Leka1996,Cheung2014}  or  shearing of coronal field by photospheric horizontal flows \citep{Klimchuk1992,Torok2003,Aulanier2005,Aulanier2010}.
%{ However the observations were uncertain due to polarisation measurement difficulties and whatsoever they could  not provide a robust interpretation why} 

{ The question of whether net currents exist of  not   have been the motivation for the development of recent MHD models, which were dedicated to this question  
\citep{Torok2014,Dalmasse2015}.}
Let us  discuss in particular  these two sets of   theoretical papers based on  mechanisms of injection of stress in the corona.\\

\citet{Torok2003}  studied the stressing  of coronal magnetic field by photospheric motions  of a bipole and noticed that net currents develop  if the  vortices are {\it close} enough to the PIL.  
Later \citet{Torok2014}, and \citet{Dalmasse2015}  used      respectively  the  Lare3D code \citep{Leake2013}  and  the { Observationally-driven High-order scheme Magnetohydrodynamic} code (OHM) \citep{Aulanier2005} to quantify  net currents and to explain why they   exist. In the  simulation  of  \citet{Torok2014} a sub-photospheric magnetic flux rope containing neutralized current was considered (Figure \ref{kevin}a).   In the case of  emergence of  a flux tube, it has been shown that the flux tube looks like  an Omega loop, starts to  flatten below the solar surface and emerges
progressively \citep{Archontis2008,Schmieder2014}.
 Unneutralized (i.e net) currents  appear   in the modeled photosphere  when the flux tube has not completely emerged. The top part with 
 direct currents
has  emerged while   a non negligible  part of  return currents surrounding the flux tube is  still below the photosphere.

 \citet{Dalmasse2015} revisited the case of   photospheric motions  inducing  stressed magnetic field by  rotating  and  shearing  the polarities of a bipole.
 In the case of  twist,  
 the photospheric vertical current density maps display   direct and strong  currents in the core of each  polarity surrounded by a shell of return currents with a swirling pattern (see Figure 3 in \citet{Dalmasse2015}). This asymmetry is due to the effect of field line length resulting from the flux tube curvature. The stronger currents develop at the footpoints of the shorter field lines. This pattern can exist only in a 3D configuration and not in 2.5  D cylindrical geometry.   It explains  on its own the whirling pattern of the  current density but not the  unneutralized currents.

  Figure \ref{kevin} is a composite figure with  panels of different models of twist and shear, all showing an important shear along the PIL which is the clue to obtaining a net current. The left column concerns an emerging flux model and  shows  the magnetic  field vectors parallel to the PIL (from \citet{Torok2014}).  In such a magnetic configuration  there is a strong shear along the PIL.  The two right columns concern  the sheared field model of \citet{Dalmasse2015}   with at the top right (e)  { for comparison}  the sketch 
  by \citet{Melrose1991} on neutralized currents
  with   lanes of positive alternated with negative currents.  This pattern is very similar, in fact, to the current density pattern (panel f) found in  \citet{Dalmasse2015}  for  a weak  shear, in which the currents  are not   neutralized. { When the PIL is sheared 
   the magnetic field does not have anymore a current free region around the PIL.}
   
 \citet{Dalmasse2015} concluded that  magnetic shear along the PIL  caused by the motions imposed in the photosphere is responsible of the unneutralized (i.e. net current) current observed in ARs and shown in   current  density maps. The  magnetic shear generates a force-free net current.  According to the different models,   unneutralized currents  occur when the  
 twist /shear  motions (i.e.  flows) reach the PIL  \citep{Dalmasse2015}. In that case, there is no way to have a potential field at  the PIL,   i.e. a  flux rope with direct current in its core surrounded by return currents. The shear inhibits the return current { along  to the PIL and only direct current   remains there. }

{   In conclusion, whatever the magnetic field geometry is, the  net current 
  depends on the  length   of the  part of the PIL  above  which the magnetic field is sheared, and also depends on how strong and how sheared the  field is. This idea is consistent with  past observations   \citep{Wheatland2000,Falconer2002,Ravindra2011},  and with the magnetic conditions found for eruptions   (\citet{Falconer2002} and Guennou et al (2017).}
  
  Several other attempts have been made to interpret 
   the  observed current density pattern. 
    %{  before } the MHD models of  \citet{Torok2014,Dalmasse2015}, proposing different mechanisms
   In particular \citet{Georgoulis2012}
  conjectured a 
  mechanism  for producing non-neutralized currents    based on a  dynamical compression which would generate a Lorentz force along the PIL.

 %{ The MHD models of  \citet{Torok2014,Dalmasse2015}  retrieved the current densities pattern  found  in recent observations after several  other attempts based on different mechanisms  \citep{Falconer2002, Ravindra2001,Wheatland2000}.  \citet{Georgoulis2012} suggested a different mechanism for producing non-neutralized  currents} based on a  dynamical compression which would generate a Lorentz force along the PIL.  

%  Several attempts have been made  to retrieve the  current density pattern   discovered in the MHD models \citep{Torok2014,Dalmasse2015} in the observations (see next sections.)

\subsection{Link between Coronal Structures and Photospheric   Electric  Currents}

In addition to the question of current neutralization, another 
  question is: can we detect flux ropes in the corona before eruptions?\\
In fact the observations of the corona in multi-wavelengths exhibit many structures that look like  flux ropes. 
 Many structures  in the corona are  considered as  direct signatures of 
 current-carrying magnetic fields such as the following:\\
 \begin{itemize}
 \item  Forward or reversed sigmoids, bright structures with a S-shape or reverse S-shape,  observed in EUV  with the Atmospheric Imaging  Assembly (AIA)  imager aboard the {\it  Solar Dynamical Observatory}  (SDO)  with the  131 \AA\, and 94 \AA\, filters { \citep{Gibson2006,Savcheva2012a,Savcheva2012b}}, and in X-ray  with Yohkoh  \citep{Canfield1999}.   Sigmoids  are frequently   { inferred to be}
 %assimilated 
   bundles of  whirling field lines forming a flux rope \citep{Dudik2014,Janvier2013,Zhao2014,Zhao2016,Cheng2011,Cheng2016}. They are considered to the  surrounding   the erupting flux rope.
  \item The prominences, called filaments when observed on the disk,  are formed by cool plasma suspended in magnetic structures with J parallel to B \citep{Aulanier2002,Mackay2010,Guo2010}. 
 Dark cavities with spinning motions    observed  at the limb in coronal lines  ({\it i.e.}  193 \AA\, and  171 \AA\,  with  SDO/AIA)  are also interpreted as being  the signatures of flux ropes \citep{Gibson2010,Parenti2012}. They are formed by persistent shear flows and flux cancellation at PILs \citep{vanballe1989}.  \end{itemize}
The intrinsic relationship between filaments,  sigmoids and magnetic flux ropes is not  obvious. 
 It is commonly very difficult to prove that the sigmoid  forms a flux rope \citep{Zhao2016}. 
Many studies interpret  sigmoids and  filaments,  as indicating the presence of a flux rope because it is  a direct interpretation for  eruptions and coronal mass ejections.
In the CME models, filaments are identified as  a tracer of the  flux rope, which can erupt due to the torus or breakout instabilities \citep{Kliem2006,Schmieder2015}.
 Flux ropes are important because they  indicate the presence of strong electric currents.  Their  footpoints in the photosphere  should be a  region with intense currents and a possible signature of their existence.

However  the detection of 
 flux ropes  in the corona and their footpoints in the photosphere is not trivial.
 In many observations, it is not clear that the filament channel  or the sigmoid plays a role in the eruption  and the eruptive sigmoid can be  quite different from it.   For example on July 12, 2012,  the filament in { AR 11520 } remains unperturbed during the entire flare. 
The flare  starts in a pre-existing coronal sigmoid which does not take part of the eruption. The eruption is occurring via   a second sigmoid formed by continuous slipping magnetic reconnection of the overlying arcades \citep{Dudik2014}. In this complicated case it is very difficult to define the flux rope and even more its footpoints  which are moving during the entire period of the eruption. 
 In many cases  sigmoid magnetic field lines  can be considered as the envelope of the flux rope. Commonly the field lines    are anchored in strong magnetic field and cannot shift easily. Therefore  sigmoids  can be used as a proxy of  a flux rope.

Nonlinear force-free field (NLFFF)   extrapolations are  also in principle a good tool for testing the existence of a flux rope.
However the magnetic field in the boundary (photosphere) has to be modified, e.g.  preprocessed  which may lead to photospheric  magnetic field with less free energy \citep{Wiegelman2010}. This is the limitation of the NLFF field extrapolation studies.
If the flux rope can be detected by a NLFFF  extrapolation, the footpoints of the flux rope would be  nevertheless better defined. { We have to keep in mind that many  different NLFFF  methods  exist and could lead to different locations of flux rope footprints. Thus the solution is also not unique. }

We present  in the following  a case of a   flux rope  anchored in a sunspot  with strong magnetic field ($\beta<1$)  in which case  the problem of the  non force-free photosphere has been  { somewhat } avoided
\citep{Zhao2016}.
Figure \ref{Sept10} illustrates the magnetic  reconstruction of the   UV sigmoid  observed on September 10, 2014 with AIA 94 \AA\,  using a NLFFF extrapolation based on the Grad-Rubin method \citep{Wheatland2013,Gilchrist2014,Zhao2016}.  One footpoint of the flux rope is surrounded by  whirling  bundles of  extrapolated field lines anchored in the leading  polarity.  The second footpoint
corresponds to the other end of the field bundles which is localized in the  weak magnetic field negative polarity. The S shape of the sigmoid  indicates that the twist of the flux rope has a negative helicity. It is a left-handed  twist  with currents anti-parallel to the magnetic field. 
The right panel in Figure \ref{Sept10} presents the smoothed 
electric current map  computed from the data of HMI,  the vector magnetograph on board SDO.   The current density $J_z$ in the positive leading  polarity has therefore a negative sign. A ring  of positive currents are well identified 
 at the edge of this  positive polarity. They correspond to the return currents. The  $J_z$  values  in the negative following polarity have  mainly a positive sign and represents the direct current density. The return current in that polarity is not detectable  presumably because of unresolved small flux tubes of the network (following  Parker, see Section 2.1).

%Figure 4
 \begin{figure*}
   \centering
    \includegraphics[width=0.8\hsize]{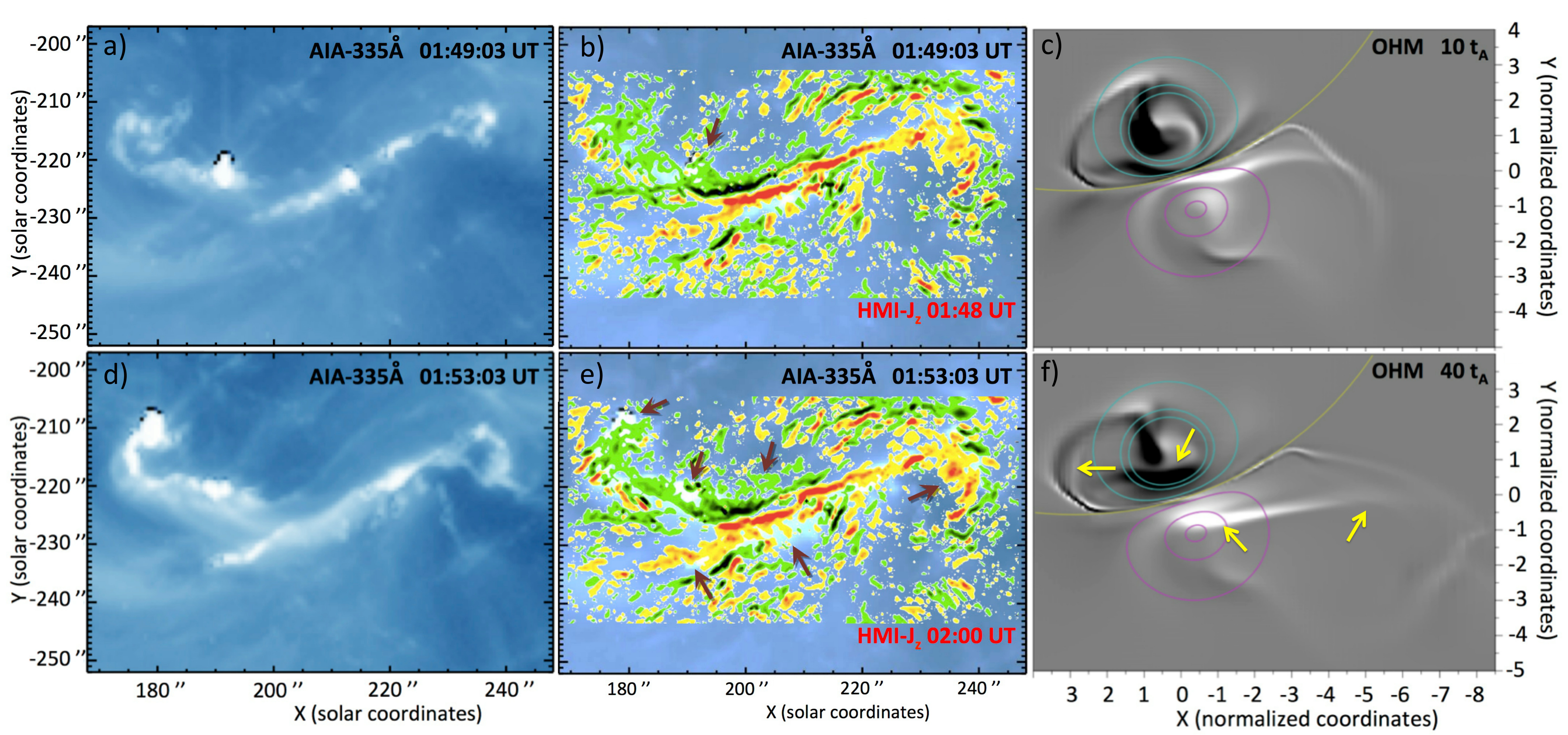}
   \caption{({\it Left column}):  Flare ribbons at the onset (a) and during (d) the peak  of the flare observed in 335\AA\, with SDO/AIA in  { AR 11158 } on February 15, 2011. 
   ({\it Middle column}): Vertical current density  $J_z$. maps  (red/green are positive/negative currents saturated to  +/- 0.2 A.m$^{-2}$) from SDO/HMI before (b) and after  (e)  the impulsive phase superposed over a 335 \AA\ image  showing the ribbons.
The  flare ribbons are very well co-aligned with the current ribbons.The brown 
 arrows show similar structures found for $J_z$ and the ribbon maps. ({\it Right column}): $J_z$ photospheric maps from OHM 3D simulations of an eruptive flare. 
 The yellow arrows point out the differences seen around the end of the simulation compared to the beginning 
  \citep[adapted from][]{Janvier2014}.}
   \label{ribbon}
   \end{figure*}

\section{ELECTRIC CURRENTS DURING ERUPTIONS}

\subsection{Current Ribbons}

 In the previous sections,  we have { discussed } now pre-eruptive ARs   are often associated   with flux rope signatures.
 We have shown an   example of  direct/return currents in current  density maps in  the center of an  active region formed by  emerging magnetic flux  before an eruption.  
 We have seen the existence of two parallel elongated current  density lanes symmetrically  (each ending with one hook) located on each side of the PIL (Figure \ref{THEMIS}).
 
 Some recent studies have also started to address the properties of active currents during solar flares. Flares  are characterized
by  a fast increase of the light emission in a wide range of the electromagnetic spectrum. During the impulsive phase the sudden  increase 
of brightening of two elongated ribbon-like structures,  which could   already exist  in the pre-phase as it has been shown previously  is    observed in visible radiation  (H$\alpha$) as well as in UV wavelengths (in C IV with SMM/UVSP in \citet{Schmieder1987}, in 304 \AA\, with TRACE in \citet{Chandra2009}, in different filters  of AIA in \citet{Dudik2014,Janvier2014}), and sometimes in hard X-ray sources \citep{Kruker2007,Musset2015}. 
The two-ribbon flare models explained the bright ribbons as due to the  impact in the chromosphere  of the accelerated particles from the reconnection site \citep{Forbes1989}.
MHD simulations obtained with the OHM code  allow  a 3D view of the phenomena. As described in the pre-phase of the flare, the elongated H$\alpha$  and UV brightenings  coincide with the photospheric footprints of the QSLs, manifested by lanes of high current density that enclose the flux rope  \citep{Janvier2013}.
Reconnection occurs in current sheets in  the quasi-separatrix layers (QSLs),   and in particular in the thin high region underneath the flux rope. The simulations predict the increase of current density in the QSLs.

We have to point out that there are very few examples of observations  of current density  maps during eruptions, up to now,  because it requires a rapid acquisition of the {\it IQUV} (Stokes parameters) in a wide region.  However the time to scan an active region requires one hour, half a hour and 12 min  with  the THEMIS, Hinode, and HMI vector magnetographs  respectively. Therefore only HMI,  { with its  relatively high cadence},  is able to follow the evolution of the current density during an eruption with a high spatial resolution. The low spectral resolution of HMI  (five  points along the line profile only)   apparently  does not  strongly influence  the determination of the currents {  because  for the inversion, the code has to fit simultaneously the four  Stokes parameter profiles for  one pixel.  So  in fact it is  20 points and not five, that are fitted simultaneously. This is still a relatively small number, but it   brings  a lot of information.
What is also important in the observations of the locations of strong  current densities  is the spatial continuity in the 2D maps. 
 For isolated current concentrations that only cover one (or a few pixels), {  the signal  could  disappear  in the noise}. 
 %it is  more difficult to reach generally  to a conclusion.}

We focus our review on  the pioneering paper  of \citet{Janvier2014},  and on the second paper \citep{Janvier2016} confirming the results found in the first paper for a  different active region. 
The first paper concerns the active region  {AR 11158  }of February 15, 2011, where a X2.1 class  flare occurred. 
 This active region was located in the northern hemisphere. This region was {  intensively studied} \citep{Inoue2015,Inoue2016,Zhao2014}.
The region consists of  the central portion of two  { emerging active regions which,  a few days earlier,  joined each other. Therefore  a strong shear  was developing between the two central polarities (leading and following) and led to X-class flares  \citep{Schrijver2011}.}
The sigmoid observed in the AIA filter by SDO, not shown here,  indicates the presence of a flux rope of positive helicity before the eruption.

Figure \ref{ribbon} focuses on the two polarities in  the AR center. In  the left panels are displayed   the ribbons, in  the middle panels,  the current density $J{_z}$ maps  obtained from HMI.
The two  bright ribbons  observed in AIA 335 \AA\  and  their  corresponding   electric current density  ribbons, $J{_z}$ (negative /positive current over the negative/positive polarity) appear  
% It looks like that  
at the onset of the flare.  The ribbons were thin and overlapped the elongated J-shaped  current density lanes observed before the flare, and  during the impulsive phase. They  very rapidly  became thicker,  because the post flare loops were already developing below.
The hooks became more
prominent   with an increase of the current
density during the impulsive phase of the flare. 
During  the eruption, the hooks increased in size. 
  Quantitative results  for  the total current {\it I} integrated in the surface of  boxes  covering the hooks showed  an  increase  by a factor reaching two \citep{Janvier2014}. In the second paper  the increase of the total current   reached an increase of 2.8 \citep{Janvier2016}.

These results have been discussed in  the frame of   two theoretical  MHD models \citep{Janvier2016}: one using  OHM  \citep{Aulanier2010}  for the first paper and  the other paper  the model of \citet{Savcheva2009}.
 Figure \ref{ribbon} (right panels)  shows  the evolution  of current density maps resulting from a 3D MHD
simulation of an eruptive flare  \citep{Aulanier2010,Janvier2014}.  
% { No attempt has been made  to analyze separately the direct and return current observationally and theoretically in these two papers.}

These  main features  are  summarized in  a sketch with the QSLs (grey-pink areas)  and  magnetic field lines  (four lines) enveloping a flux rope (Figure \ref{sketch}). The QSls  are coronal
current layers, that extend   to the photosphere  and  are  the locations of high
electric current densities. The magnetic field lines are distorted at  the QSLs  which are a   possible site of reconnection, as  indicated by red arrows.
The footprints of the 
QSLs and current layers are J-shaped (Figure \ref{ribbon} right panels) and the hooks
surround the legs of the flux rope. The magnetic topology  analysis of the February 15 2011  active region confirmed that the bright ribbons  of the flare (observed in AIA/304 \AA) overlaid the QSL footprints in the photosphere (see Figure 4 in \citet{Zhao2014}). The QSL  footprints   wrapped around  the flux rope footpoints in the photosphere.

The  numerical simulations  show that, in  torus-instability models, the photospheric current ribbons can be  interpreted { qualitatively} as the footprints of the 3D coronal current following the QSLs \citep{Janvier2014,Janvier2016}.   The simulations being running  dimensionless,  no quantitative analysis has been performed on these simulations yet.
%The evolution of the current density has been  studied by means of  MHD  numerical simulations. 
 As the torus instability develops,  { stronger  electric currents } are formed along the QSLs and  reconnection is not fast enough to destroy them,   as it is limited by the Alfv\'en speed \citep{Lin2001}.  The first observations have confirmed  that the total current  increased by a factor around two during the flare. 
 %{ No  quantitative investigation on the  total current  evolution  has been done in the MHD  simulations yet.}
  
   %Figure 5
\begin{figure}
   \centering
    \includegraphics[width=7cm]{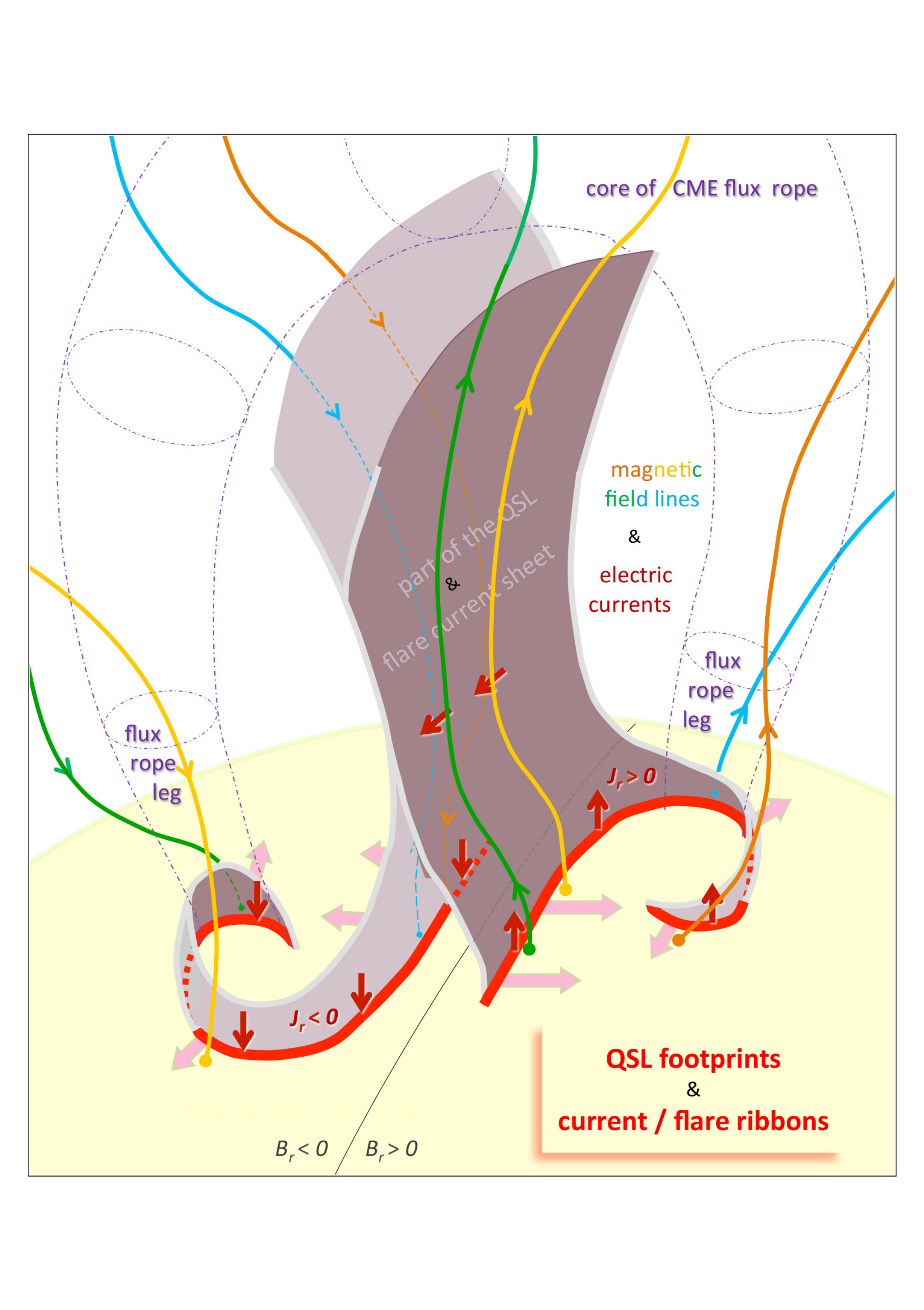}
   \caption{Cartoon for the standard model in 3D with four pre-reconnected
magnetic field lines (yellow, green). The  pink-grey area represents
parts of the 3D volume of the QSLs and the current layers. The red lines with hooks represent the footprints of the QSLs in the photosphere and are similar to two ribbon flares.
   The dashed lines with oval sections in the overlaying area of the QSLs represent the flux rope with its hook footprints in the photosphere
    \citep[adapted from][]{Janvier2014}.}
   \label{sketch}
   \end{figure}

%Figure 6 
 \begin{figure*}
   \centering
    \includegraphics[width=\hsize]{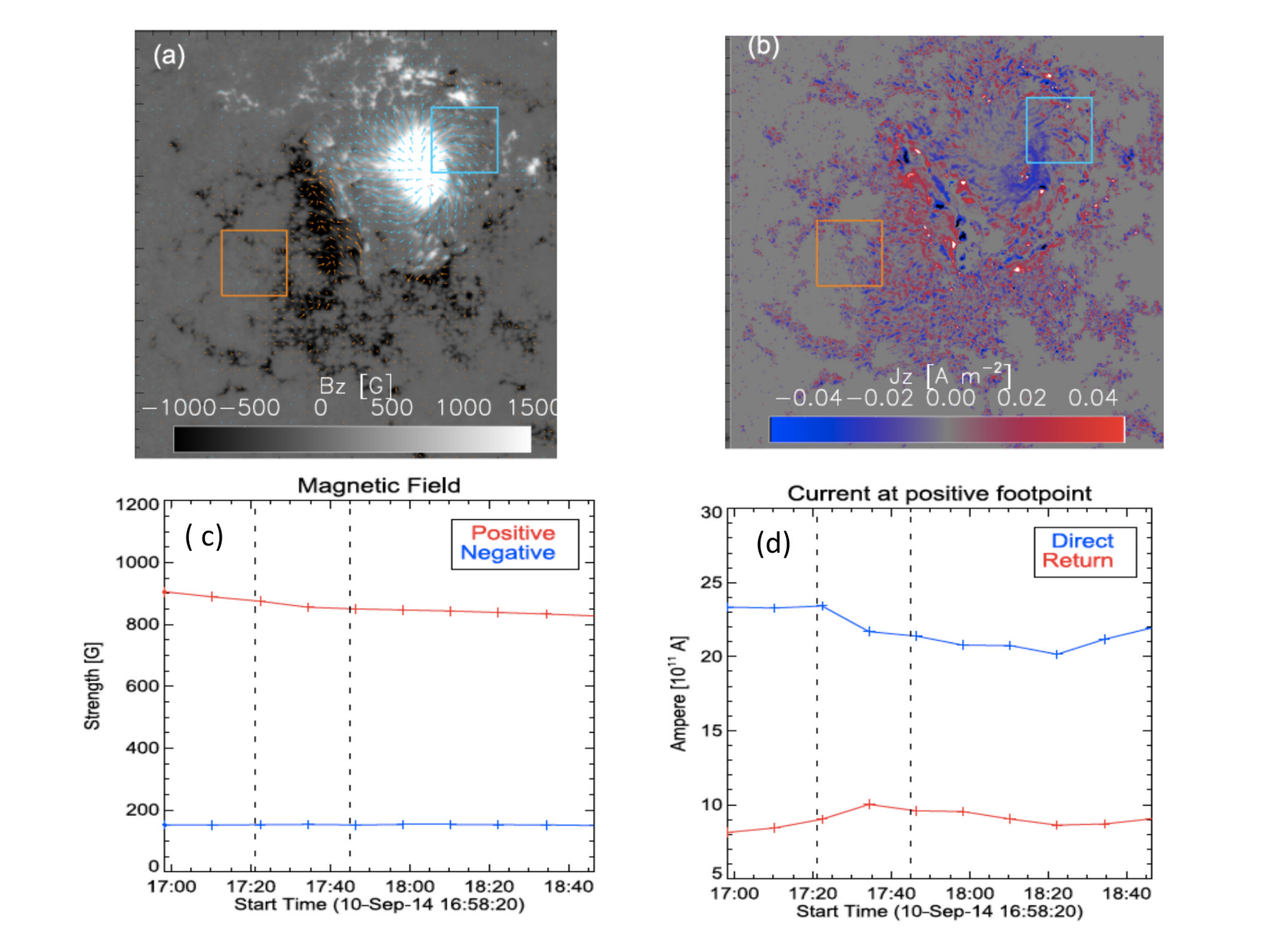}
   \caption{Photospheric electric currents in the footpoints of  the sigmoid/flux rope in  AR  12158 on September 10, 2014 (the same AR as in Figure \ref{Sept10}). (a) HMI vector magnetogram with  two boxes surrounding the deduced  two  flux rope footpoints; (b) HMI electric current maps; ( c)  the  mean strength of the magnetic field measured in the two boxes of panels (a) and (b);  (d)   the direct and return current in the positive magnetic polarity box. The two vertical lines indicate the onset and peak of the flare  (adapted from  \citet{Cheng2016}).}
   \label{Cheng}
   \end{figure*}
\subsection{Current Decrease at the Footpoints of the Erupting  Flux Tube During and After the Eruption}

As  explained in the previous section, the prediction of MHD models is that 
the localized  electric currents increase inside the flare ribbons which are the intersections of the QSLs and the photosphere \citep{Aulanier2012}.  
On the other hand, the MHD model predicts also that the electric current inside the flux rope should decrease  due to the expansion of the flux rope  \citep{Aulanier2005}.  This may be  deduced from flux conservation. It was demonstrated  that for a cylindrical  flux rope with a constant end-to-end twist, the current $J_z$  is proportional  to $L^{-1}$, $L$ being its length (see equations (46) and (47) in  \citet{Aulanier2005}). During the eruption the flux rope is stretched so  its length increases,  and  the footprint currents in the photosphere should decrease. The current decreases also in catastrophic models \citep{Forbes2000}. 

%This second prediction
{ However the decrease of the electric current inside the flux rope}  is difficult to test because of the uncertainty of the identification of the flux rope ends (see Section 2.2). However, combining two recent papers studying the same active region AR 12158 on September 10, 2014, we can conjecture that the ends of the sigmoidal loops wrapping around  the flux rope  found by extrapolation by \citet{Zhao2016}   could  correspond to the ends of the sigmoid observed in the  hot channel of AIA (94 \AA) and used as proxy of the flux rope  ends in the \citet{Cheng2016} paper (see Figures  \ref{Sept10} and \ref{Cheng}).
\citet{Cheng2016}  studied  the evolution of the magnetic field and the currents in  two boxes contained the footpoints of the sigmoid. 
One  flux rope footpoint is  located in a strong  field  positive polarity.  The second   one, being in a weaker negative  field  corresponds to  weak currents,  which could be at the limit of the measurements.  
A decrease of the direct current is measured in the strong positive polarity (Figure \ref{Cheng}d).
 These results agree with the qualitative behavior of numerical simulations  which treat  asymmetrical configurations,  similar  to the observed configuration of this AR  \citep{Aulanier2010}. 

{ The decrease of the currents is predicted  by   MHD models because of end-to-end twist conservation induced by line tying  \citep{Aulanier2005}
and catastrophic models  because of flux conservation below an erupting flux rope \citep{Demoulin2010}. This is opposite to the behavior that is built in circuit models \citep{Melrose1991} where the total current is prescribed.   Direct and quantitative comparisons between models and observations are therefore needed to better
characterized the time-evolution of currents during solar flares.}

\section{CONCLUSION}

This chapter  discusses    electric currents in the pre-eruption state and
in the course of eruptions of solar magnetic structures, using information
from solar observations, nonlinear force-free (NLFF) field extrapolations
relying on these observations and three-dimensional magnetohydrodynamical
(MHD) models.
With the new generation of vector magnetographs new current density maps have been obtained showing in  each  polarity  a complex  whirling pattern of current density \citep{Schmieder2012}.

{ The main topic of this chapter is to show 
how the new current density maps   in active regions combined with the development of new  3D MHD models   allows us  to   progressively  interpret the observations  and better fine-tune the understanding of MHD mechanisms, in the context  of eruptions.
MHD  simulations of pre-eruptive active regions,    based on the  coronal flux rope concept.
% separated from overlying arcades by quasi-separatrix layers (QSLs), 
show  a whirling electric current  density pattern, similar to the observations  in the photosphere with direct and return current in each polarity of the initial bipole \citep{Aulanier2005,Aulanier2010,Janvier2014}.  The current density  occurrence  in bright flare ribbons in the footprints of the QSLs  that separate the flux rope from the overlying arcades  was an important discovery.

Before  discussing eruptions, we have reported how the old problem of unneutralized currents has been resolved. The new  observations of electric current density  of pre-eruptive active regions have reopened the long debate on the neutralization of the currents  in the ''90s" by \citet{Melrose1991} and \citet{Parker96enddebate}.   
The key parameter  determining   the existence of  unneutralized (i.e. net current) active regions in the pre-eruptive phase  has been proven to be the  shear along the PIL of a bipole \citep{Torok2014,Dalmasse2015}. 
 Unneutralized photospheric current density patterns appear whenever magnetic shear 
is present along polarity  inversion lines (which is typical of observed intense PIL,  e.g. in  sunspots). { 
In the models of \citet{Dalmasse2015}, shear and twist were created in the corona by artificial line-tied surface motions that were primarily aimed at building a series of differently sheared quasi force-free configurations. In reality shear could be due to  }
%The shear is caused  by 
the coupling of the dynamics of the emergence of  a flux rope and the convection in the sub-photosphere. This shear  creates the net current { when  the flux rope does not emerge completely}. Shear  along the PIL and  net current are  necessary for the occurrence of  flares and eruptions  \citep{Falconer2002,Kliem2006} .

A whirling current density pattern, observed in some active regions before eruptions,  has been interpreted as the photospheric  footprint of flux ropes in the corona. The currents have a hook shape which, according to the models encircles the flux rope footprint.  Inside the active region two elongated current  density lanes are detected,  symmetrically located along the PIL  when there is a strong shear before  the flare.
All  regions do not show such a structure before the eruption but already a few cases have been published \citep{Schmieder2012,Janvier2014,Georgoulis2012}. 
In the observed current density pattern   with J-shaped   hooks, the   high  spatial resolution  allow us  to distinguish  the currents in the hooks and the surrounding     mixed sign currents.  Considering one  magnetic polarity, i.e. above a sunspot,  it is clear that both direct and return currents are present. The direct current is generally dominant.
During the eruptions  more intense current ribbons  are observed  and an increase of the current density in the straight part of the ribbons  has been  measured. 
The  total current  during an eruption in the current ribbons  including their   hooks    has  been quantified   in two observational cases up to now, and shows an increase by a factor of two \citep{Janvier2014,Janvier2016}.
This can be explained qualitatively.
When a flare begins, new narrow J-shaped current structures develop on top of the pre-existing relatively broad  elongated current density lanes, also in J-shape. These new structures appear to match bright flare ribbons, visible from the visible to EUV wavelengths.
On the contrary  in regions encircled by the hooks a decrease  has been detected \citep{Cheng2016}.  This can be explained by 
 the expansion of the flux rope. 

 Several authors  analyzed  MHD simulations   before and during an eruption  from a qualitatively point of view \citep{Torok2014,Janvier2014,Dalmasse2015}. 
 The simulations { use dimensionless units} and  have not yet analyzed  to give quantified results on the values  of the current increase during an eruption, nor  how much  of a is decrease in the current density in the flux rope we  can expect.    No attempt of  distinguishing direct and return current in ARs during eruptions has been carried out yet. The work still remains to be done. With the increase of  computer facilities and the parallelization  of codes it should be possible to estimate the net current in each polarity and to compute their variation during eruptions.
 
Today we have vector magnetograms with high spectral resolution (Hinode/SOT, THEMIS)  in a small field of view  with low time resolution. We have HMI full disk vector magnetograms with  high cadence but relatively low sensitivity.
 In the future we may expect  to have 
 sensitive vector magnetograms with  higher  spatial resolution    to compute  electric current density  with accuracy and with higher  cadence to study the evolution during  eruptions.}

 DKIST and EST  (the  future  4 meter telescopes) and Solar Orbiter/PHI { (the Polarimetric and Helioseismic Imager)}, which will fly close to the Sun,  will  certainly bring very surprising  results in this domain, on small  field of views with DKIST and EST  {  and  on the full disk with PHI.}

\bibliographystyle{aa}
\bibliography{CurrentsXflare.bib}

\end{document}